\documentclass[10pt, format=manuscript,nonacm]{acmart}

\usepackage{booktabs}
\usepackage{tabularx}
\usepackage{xcolor,colortbl}
\usepackage{tabularx}
\definecolor{lightgray}{gray}{0.93}
\definecolor{slightgray}{gray}{0.98}
\definecolor{darkgray}{gray}{0.77}
\usepackage{makecell}
\usepackage{multirow}
\usepackage{pgf} 
\usepackage{threeparttable}

\usepackage{xparse}   

\NewDocumentCommand{\calc}{mm}{
  \pgfmathparse{#1/#2*100}\pgfmathprintnumber[fixed,precision=2]{\pgfmathresult}
}

\AtBeginDocument{%
  }

\setcopyright{acmlicensed}
\copyrightyear{2018}
\acmYear{2018}
\acmDOI{XXXXXXX.XXXXXXX}

\acmConference[Conference acronym 'XX]{Make sure to enter the correct
  conference title from your rights confirmation emai}{June 03--05,
  2018}{Woodstock, NY}
\acmISBN{978-1-4503-XXXX-X/18/06}

\begin{document}
\title{Understanding Design Fixation in Generative AI}
\renewcommand{\shorttitle}{Design Fixation in GenAI}

\author{Liuqing Chen}
\email{chenlq@zju.edu.cn}
\affiliation{%
  \institution{College of Computer Science and Technology, Zhejiang University}
  \city{Hangzhou}
  \country{China}
}
\author{Yaxuan Song}
\email{songyx23@zju.edu.cn}
\affiliation{%
  \institution{College of Computer Science and Technology, Zhejiang University}
  \city{Hangzhou}
  \country{China}
}

\author{Chunyuan Zheng}
\email{3200103594@zju.edu.cn}
\affiliation{%
  \institution{College of Computer Science and Technology, Zhejiang University}
  \city{Hangzhou}
  \country{China}
}

\author{Qianzhi Jing}
\affiliation{%
  \institution{College of Computer Science and Technology, Zhejiang University}
  \city{Hangzhou}
  \country{China}
}

\author{Preben Hansen}
\affiliation{%
  \institution{Stockholm University}
  \city{Stockholm}
  \country{Sweden}
}

\author{Lingyun Sun}
\affiliation{%
  \institution{International Design Institute, Zhejiang University}
  \city{Hangzhou}
  \country{China}
}

\renewcommand{\shortauthors}{Trovato et al.}

\begin{abstract}
Generative AI (GenAI) provides new opportunities for creativity support, but the phenomenon of GenAI design fixation remains underexplored. While human design fixation typically constrains ideas to familiar or existing solutions, our findings reveal that GenAI similarly experience design fixation, limiting its ability to generate novel and diverse design outcomes. To advance understanding of GenAI design fixation, we propose a theoretical framework includes the definition, causes, manifestations, and impacts of GenAI design fixation for creative design. We also conducted an experimental study to investigate the characteristics of GenAI design fixation in practice. We summarize how GenAI design fixation manifests in text generation model and image generation model respectively. Furthermore, we propose methods for mitigating GenAI design fixation for creativity support tool design. We recommend adopting the lens of GenAI design fixation for creativity-oriented HCI research, as the unique perspectives and insights it provides.

\end{abstract}




\maketitle

\section{Introduction}

In the evolving landscape of human-computer interaction (HCI), Generative AI (GenAI)-enhanced tools have emerged as pivotal in advancing design creativity support research \cite{oh2024lumimood, choi2024creativeconnect, lin2024jigsaw}. Central to the appeal of GenAI in design is its adeptness at synthesizing contextual solutions. This strength stems from the integration of large datasets, algorithmic architectures, and users' prompt \cite{schellaert2023your}. However, these same properties also impose constraints on guiding GenAI systems to be creative, with documented limitations in generating truly novel and diverse design outcomes, particularly due to a tendency towards homogenized design perspectives \cite{doshi2024generative, kobiella2024if}.

Recent studies have begun to raise the question about the extent to which these models can produce genuinely creative outputs, since the data used to train models are highly-centralized, data-driven. Some recent user studies of Large Language Models (LLMs) provide evidence of homogenization effect in LLM-supported writing \cite{padmakumar2023does, doshi2023generative}. One such study found that the application of GenAI reduced the collective diversity of novel content \cite{doshi2024generative}. However, the authors of these articles did not deeply analyze the potential causes of this phenomenon, instead, they mainly talked about the homogenization effects on human creative ideation \cite{anderson2024homogenization, doshi2024generative}. Indeed, recent work has called for foundational research to understand what constrains GenAI’s output, and conform their existence outside of writing tasks specifically. Here, we argue that, similar to design fixation as an inherent cognitive pattern in the human design process, Generative AI also exhibits design fixation. Firstly, we suggest that current GenAI systems exhibit design fixation; understanding this phenomenon can help interpret some previous research and reveal challenges that limit the novelty and diversity of GenAI outputs. 

While the concept of fixation is well-recognized in the HCI community, research has primarily focused on developing tools to mitigate human design fixation, often discussing design goals or benefits of such systems \cite{yoo2024bi, lee2024conversational, chung2022artistic, lamiroy2022lamuse}. Another few research towards the examination of design fixation focusing on human side \cite{wadinambiarachchi2024effects}. However, there is a lack of systematic and rigorous investigation of GenAI design fixation. This oversight of the constraint of GenAI capabilities may obscure certain issues in creative support that are otherwise overlooked by other lens for design creativity support. \cite{li2023beyond, frich2019mapping, frich2018twenty}.



While highlighting the advancements, discussions about flaws or imperfections within the outputs in creativity support are also becoming necessary to enhance scientific balance \cite{weisz2024design}. Reflections and examinations about GenAI creativity have arisen, which mainly focus on creative writing area. For example, \cite{chakrabarty2024art} evaluating the creativity of LLM-generated stories compared to professional authors. \cite{doshi2024generative} examines the relationship between GenAI and human creativity, which demonstrates the narrower scope of novel content based on GenAI production. Meanwhile, GenAI still remains elusive. It derives opaque processing models from data patterns, and its outputs are inherently uncertain due to generative variability \cite{weisz2023toward, weisz2024design}. Generative AI's imperfection also draws the attention of HCI community, such as hallucination, AI error and bias. \cite{benjamin2021machine} proposed that AI "error" could serve as an unexpected material to help designers explore new spaces for design intervention. \cite{van2022ceci} found that AI "error" could bring design reflection through surprise. \cite{liu2024smart} summarize the design dimensions and implementation methods of AI error for creative design.


With the growing advancement and application of GenAI, the call for a fixation lens in HCI has become urgent and important. Firstly, the generative variability of GenAI brings the risk of unconscious bias. As previous creativity support methods are based on specific vocabulary-based command or data-driven methods such as semantic networks \cite{luo2019computer, shi2017data} or knowledge graphs are generally based on explicit construction algorithms, resulting in controllable and explainable recommendations. In contrast, generative models trained on large amount of datasets may incorporate unconscious bias present in the training data. Secondly, in terms of the human-computer collaboration process, traditional creative support tools primarily serve an auxiliary role, relying on humans to drive the solutions tailor to the current problem. However, GenAI could generate "situation-based" stimuli allows for the generation of complete creative outcomes with minimal or no direct human intervention \cite{zhu2023generative}. Recent study have found that this led to an over-reliance by human users on the results generated by AI, resulting in fixation on solutions produced by GenAI \cite{wadinambiarachchi2024effects}.

Towards this aim, this study proposes a previously overlooked phenomenon—GenAI design fixation. Firstly, we define the definition and analyze the causes of GenAI design fixation. Through experiments, we investigate whether novice designer can perceive the existence of GenAI design fixation, and we summarize the impact of GenAI design fixation on the human-GenAI co-ideation process, collecting empirical evidence of GenAI design fixation in the human-GenAI collaboration process. Subsequently, we propose measures to mitigate GenAI design fixation, providing references for future Generative AI-based creative support tools. Finally, we summarize the distinctions between GenAI design fixation and other GenAI flaws and imperfections, especially in the realm of creativity, and propose possible future directions for GenAI design fixation.

In summary, we have made the following contributions to the HCI community:
\begin{enumerate}
    \item \textbf{Theoretical framework of GenAI design fixation: }We have developed a preliminary theoretical framework for GenAI design fixation by combining theoretical and empirical approaches. This framework includes the definition of GenAI design fixation, its causes, manifestations, impact on creative design, and measures to mitigate it.
    \item \textbf{Design strategies for GenAI-based creative support tools: }We use the GenAI design fixation lens to identify the needs and provide concrete directions for further research into overcoming the design fixation challenges of GenAI.
\end{enumerate}

In the next sections, we define GenAI design fixation, comparing the differences and correlations between GenAI design fixation and human design fixation in Section~\ref{section_definition}; summarize how design fixation manifest in text generation and image generation models respectively through an empirical experiment, analyzing the impact of GenAI design fixation on the creative process in Section~\ref{section_experiment}, and propose ways to mitigate GenAI design fixation in Section~\ref{section_solution}.
	

\section{Conceptualizing GenAI design fixation}
\label{section_definition}
Our preliminary definition of GenAI design fixation builds on \cite{crilly2017next}'s description of design fixation "Design fixation is a state in which someone engaged in a design task undertakes a restricted exploration of the design space due to an unconscious bias resulting from prior experiences, knowledge, or assumptions”. 


In this study, we define and explore the phenomenon of Generative AI design fixation by drawing parallels to established
concepts of design fixation as: \textit{GenAI design fixation is the state in which a Generative AI model restricts its design exploration of the generative space due to unconscious bias stemming from technical aspects and human factors, which limits the diversity and originality of the model's design output, leading to repetitive or constrained results}.

\subsection{Preliminary causes analysis}

Our investigation into the causes of GenAI design fixation began with a comprehensive review of recent literature within the HCI and AI communities. We conducted a thorough search of key research studies, creativity support tools, and technical papers pertinent to Generative AI, utilizing resources such as the ACM Digital Library and Google Scholar. Search terms included "Generative AI", "design support", and "creativity support". To enhance our understanding of the cognitive differences between GenAI and human designers, we also explored literature on design science and design cognition. This review helped identify a corpus of relevant publications, including some recent studies that discuss the limitations of GenAI in creativity support \cite{doshi2024generative, kobiella2024if, anderson2024evaluating}. The detailed review process and results are presented in Appendix A in our supplementary materials.

Our analysis suggests that the causes of GenAI design fixation extend beyond the inherent limitations of the models; human factors also play a crucial role. Thus, we initially categorize these causes into technical aspects and human factors. In the following list, the technical aspects are specifically tailored to address the context of GenAI design fixation, while the human factors illuminate the drawbacks summarized from previous literature during the human-GenAI interaction process. A more detailed analysis of these causes will be presented in Section~\ref{section_experiment}.

\begin{enumerate}
    \item \textbf{Technical aspects}
    \begin{enumerate}
        \item \textbf{Data limitation:} Data bias arise from the dataset being unbalanced or imbued with prejudiced information \cite{manduchi2024challenges}, leading to a tendency in models to learn and perpetuate these bias. Such bias can cause the model’s outputs to lack diversity in style, content, or concept, thus influencing the range of generated solutions \cite{tan2020improving, jahanian2019steerability}.
        \item \textbf{Architecture limitation:} The architecture of a model, such as the Transformer model used in text generation \cite{vaswani2017attention} or the diffusion model employed in image generation \cite{ho2020denoising}, fundamentally shapes how inputs are processed and outputs are generated. Each architecture comes with inherent constraints that potentially predispose the models to certain types of outputs, thereby affecting the generative diversity. 
    \end{enumerate}
    \item \textbf{Human factors}
    \begin{enumerate}
        \item \textbf{Misalignment in design problem definition:} During the design ideation stage, there exists a fundamental misalignment between the requirements and design goals as understood by human designers compared to GenAI systems. This discrepancy can lead to designs that do not fully meet the intended needs or expectations.
        \item \textbf{Challenges in adhering to designerly thinking:} GenAI often struggles to produce outcomes that align with the designerly way of thinking \cite{chen2024designfusion}, which emphasizes creativity, user-centered solutions, and iterative refinement \cite{zhou2024understanding}. The AI's output may not effectively reflect these nuanced aspects of design thinking, limiting the innovation potential.
        \item \textbf{Complexities of prompt writing:} Writing effective prompts for GenAI requires a deep understanding of both the design problem and the AI's capabilities. The complexity of crafting such prompts can be a significant barrier, as poorly formulated prompts may lead to irrelevant or suboptimal design outputs \cite{brown2020language, wu2022ai}.
    \end{enumerate}
\end{enumerate}


\subsection{Distinctions and correlations between GenAI design fixation and human design fixation}
In discussing the distinctions and correlations between GenAI design fixation and human design fixation, we follow the framework in a design fixation review research \cite{alipour2018review}, which propose fundamental factors regarding design fixation research including source, design process and design outcomes. 


Unlike humans, who are susceptible to pre-conceived ideas and concepts, GenAI design fixation predominantly stems from imbalances within the dataset. During the information processing phase, humans are influenced by three cognitive phenomena \cite{jansson1991design}: "functional fixedness," which refers to the challenge in perceiving objects beyond their conventional uses \cite{duncker1945problem}; "mental set," denoting a reluctance to deviate from known strategies \cite{luchins1959rigidity}; and "the path of least resistance," indicative of the tendency to minimize effort in creative tasks \cite{ward1994structured}. In contrast, the algorithmic architecture of GenAI dictates input processing and harbors intrinsic constraints that similarly restrict creative potential, leading to less innovative solutions.

In the manifestation of design fixation, human designers tend to adhere to pre-conceived ideas and concepts, which limits the exploration of the design space during ideation. Previous findings show that this is reflected distinctly in the similar characteristics between the example solutions for design tasks and the final design outcomes. Similarly, GenAI design fixation, while capable of generating targeted solutions, remains confined within the bounds set by its training data, which prevents the models from maximizing creative potential and leads to unoriginal design solutions.

\begin{figure*}[htp]
    \centering
    \includegraphics[width=0.95\textwidth]{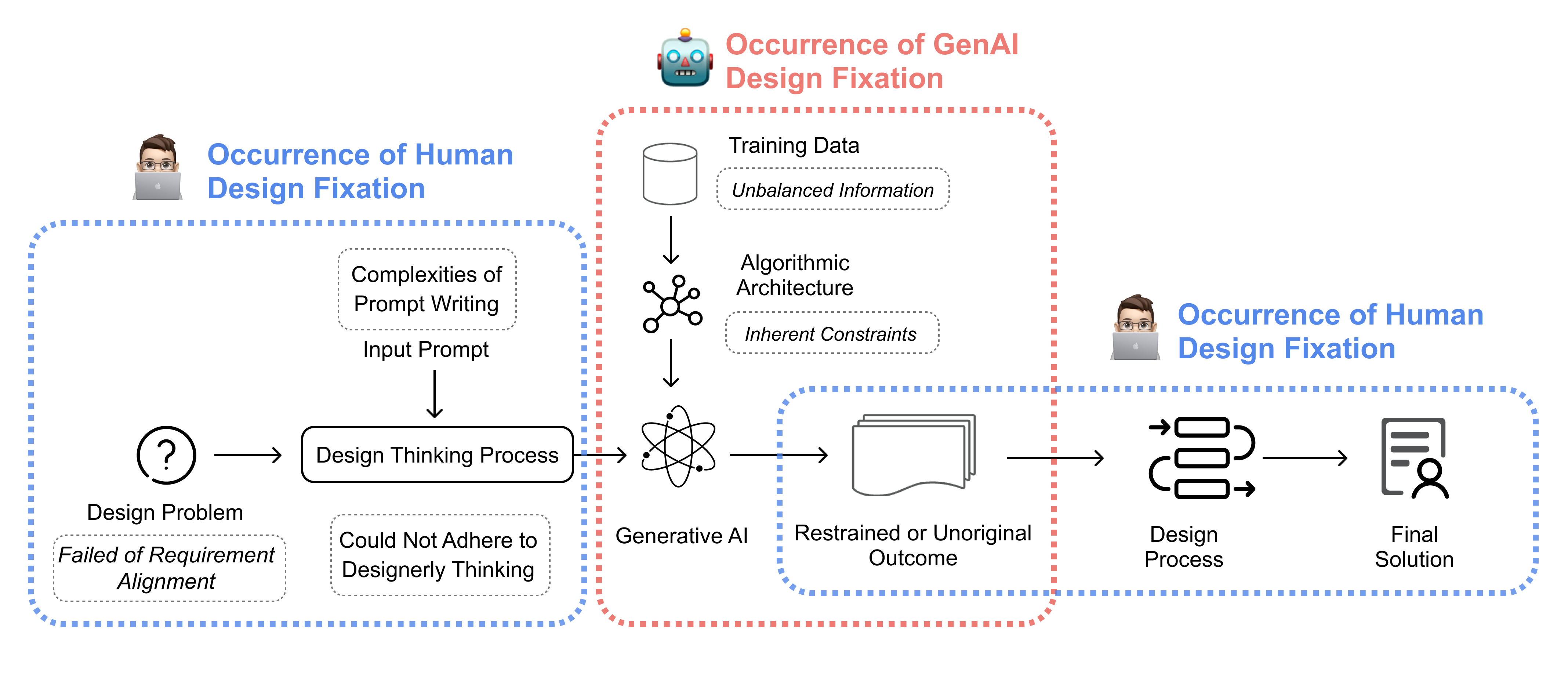}
    \caption{The process and dynamics of fixation within GenAI systems and how it correlates with human fixation in design processes.}
    \label{fig:process}
    \Description{}
\end{figure*}

Figure~\ref{fig:process} outlines the process and dynamics of fixation within GenAI systems and how it correlates with human fixation in design processes. The flowchart details how both types of fixation move from the initial design problem through various influencing factors like input prompts and training data (for GenAI) or inspiration sources (for humans) to the generated solution or final design outcome.

\section{Understanding GenAI design fixation in practice}
\label{section_experiment}
\subsection{The aim of this study}
As we have proposed the definition of GenAI design fixation in Section~\ref{section_definition}, the aim of this experimental study is to gain a practical understanding of the performances of GenAI design fixation within the human-GenAI co-ideation process. Given that this is the first instance where the lens of GenAI design fixation is applied to examine the human-GenAI co-creation process, and considering the distinct research questions concerning fixation and creativity, we conduct an experimental study to gather empirical evidence on GenAI design fixation and users' perspectives on this phenomenon. Our expected results are as follows:
\begin{itemize}
    \item \textbf{RQ1: What are the manifestations of GenAI design fixation in text generation and image generation models?}
    \item \textbf{RQ2: How do participants identify and describe design fixation in Generative AI?}
    \item \textbf{RQ3: What impacts does GenAI design fixation have on the ideation process during human-GenAI collaboration?}
\end{itemize}



\subsection{Participants}

Participants were recruited for this study through social media and word-of-mouth, targeting novice designers with experience in product design and an interest in the application of Generative AI in design fields. We focused on novice designers because they represent a demographic that is still forming their design habits and are likely more open to integrating new technologies such as Generative AI into their workflow. This recruitment strategy ensured that all participants have not only a basic understanding of the methods interacting with GenAI, but also a keen interest in sharing their experience and perspectives on ideation processes facilitated by GenAI. Ultimately, ten participants were successfully recruited, consisting of four males and six females aged between 20 and 26 years (Mean = 22.7, SD = 3.01). They come from diverse academic backgrounds, including undergraduate, master’s, and doctoral students. Additionally, one participant works in a GenAI practice organization. Participants received a commemorative gift as a token of appreciation for completing the study tasks.

\subsection{Procedure}
\label{experiment_procedure}
\begin{figure*}[htp]
    \centering
    \includegraphics[width=0.95\textwidth]{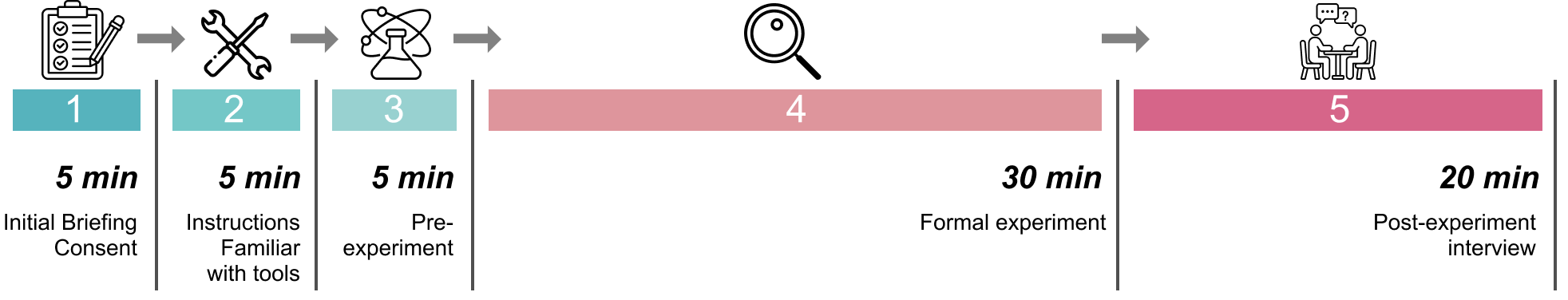}
    \caption{The process of participants engaging in our experiment.}
    \vspace{-0.05in}
    \label{fig:experiment}
    \Description{}
\end{figure*}

In this study, two researchers conducted one-on-one sessions with participants to observe each individual’s interaction with Generative AI, focusing on their ability to recognize the presence of GenAI design fixation. The scheduling of these experimental sessions was based on the availability of the participants. Participants were compensated for their time. The experimental process that each participant underwent is detailed in Figure \ref{fig:experiment}.

At the start of the study, researchers explained the experimental procedure, gathered informed consent, and collected demographic information from the participants (as shown in Table~\ref{tab:participants}). Then, participants were briefed on the design task which involved creating as innovative a design as possible under the assistance of two different Generative AI tools. Additionally, we introduced CombinatorX, a representative GenAI-assisted design method \cite{chen2024foundation}, which utilizes combinational creativity—merging two concepts into a novel idea \cite{boden2004creative}. The CombinatorX process involves identifying additive embodiments and forming textual combinational ideas, followed by using text-to-image technology to visualize these ideas. It is necessary to clarify that while the CombinatorX method was provided as a reference, its use was not mandatory but offered as a scaffold for participants lacking inspiration.

In our study, designers were instructed to produce as many design proposals as possible within the allotted time, ultimately submitting all their satisfactory designs. Each submission consisted of design descriptions generated by ChatGPT (GPT-4o), images produced by Midjourney, and the accompanying prompt descriptions. Pre-experiment phase allowed participants to familiarize themselves with the use of ChatGPT, Midjourney, and our provided ideation reference method, CombinatorX. 

The formal experiment involved a specific design task requiring participants to create innovative chair designs for office settings. As chair design involves both functional and aesthetic elements that can greatly benefit from the innovative possibilities offered by AI, the experiment focused on office scenarios to refine the creative direction and functional requirements of the designs. By focusing the chair design task on office scenarios, the experiment was structured to refine the creative direction and functional requirements of the designs, making our subsequent interviews and analysis more focused. The experimental process, including the procedure details and time limits, was established based on our pilot study. 

After the experiment concluded, we conducted semi-structured interviews, each lasting approximately 20 minutes. These interviews were informed by our observations of the participants’ design processes and the outputs generated by GenAI. We structured our interview around three topics to understand participants' submitted solutions under the help of GenAI (question 1-2), designers' creation processes (question 3-4), and designers' perspective on the output (question 5).

\begin{enumerate}
    \item ``Have you noticed any repetition or similarities in the ideas or designs generated by the tools?'' (\textbf{Corresponding to RQ2})
    \item ``Do you think there are stylistic or paradigmatic limitations in the ideas generated by these tools compared to those generated through traditional design methods?'' (\textbf{Corresponding to RQ3})
    \item ``If the design process required further development, how do you think the performance of these tools would affect the originality or diversity of your final creative outputs?'' (\textbf{Corresponding to RQ3})
    \item ``What do you think are the reasons for any limitations you perceived in the model outputs? Are these limitations due to the model itself or the interaction process?'' (\textbf{Corresponding to RQ2})
    \item ``Could you share your views on the design approaches demonstrated during the experiment, including both pros and cons?'' (\textbf{Corresponding to RQ3})
\end{enumerate}

These questions were designed to explore various aspects of GenAI design fixation. While it was found in participants' responses that several participants were already familiar with the concept of design fixation, we deliberately avoided using terms like “GenAI design fixation” or “design fixation” directly in our interviews. Instead, we used alternative descriptions  (i.e. interview question 1 above) to probe for any sensations of fixation that may have occurred during their interactions with generative AI tools. At the conclusion of the experiment, we explained the study’s aims and the terminology used to all participants.

\begin{table}[ht]
\centering
\caption{Participant demographics in our experiment.}
\label{tab:participants}
\begin{tabular}{@{}cccccc@{}}
\toprule
\textbf{ID} & \textbf{Age} & \textbf{Gender} & \textbf{Design experience} & \textbf{Use of text generation models} & \textbf{Use of image generation models} \\ \midrule
P1 & 22 & Male & less than one year & Used before & Used before\\
P2 & 22 & Female & three years & Proficient & Used before \\
P3 & 23 & Female & One year & Proficient & Used before \\
P4 & 23 & Female & Five years & Proficient & Used before \\
P5 & 24 & Female & Five years & Proficient & Proficient \\
P6 & 20 & Male & Two years & Proficient & Used before \\
P7 & 20 & Male & One year & Proficient & Used before \\
P8 & 24 & Male & Six years & Proficient & Used before \\
P9 & 23 & Female & Six years & Used before & Used before \\
P10 & 26 & Female & Five years & Proficient & Used before \\ \bottomrule
\end{tabular}
\end{table}

\subsection{Data collection and analysis}
\subsubsection{Data collection}
We collected two sets of data. The first set of data consists of the GenAI design solutions collected in Section~\ref{experiment_procedure} (totaling 96). The second set of data comes from the chair design examples from the Red Dot design contest\footnote{\url{https://www.red-dot.org/}}. The reason why this dataset were collected is for the assessment of the originality and creativity of GenAI-generated chair designs in an office setting. We choose a design contest due to its diversity and the high quality of design innovation it represents. Commercial dataset such as IKEA were avoided due to potential stylistic uniformity linked to specific brands, which could skew the diversity of the dataset. Specifically, we crawled and filtered 105 chair design entries with "office chair" containing in their design description. For these selected cases, both the images and textual descriptions of the designs were collected. This dataset provides a comprehensive foundation for comparing traditional and AI-generated designs, enhancing the analysis of GenAI's impact on design creativity and fixation in real-world scenarios.

The reason why we did not conduct between-subject analysis is that it is difficult for human designers to produce photo-like chair design schema in a limited time in a laboratory experiment.  Additionally, our research aim is to assess the novelty and diversity of GenAI-generated design outcomes, making the use of recognized design award entries as a baseline an acceptable approach for comparison.

\subsubsection{Data analysis}
\paragraph{\textbf{Text generation analysis}}
In the data preprocessing phase, to enhance the relevance analysis between descriptive content and chair design, we initially reviewed all the solutions and annotated terms specifically related to chair design such as 'seat', 'leg', 'backrest', etc. (The detailed list is provided in Appendix B in our supplementary materials). We then filtered out words representing structural design. These terms serve as indicators of the degree of alignment between the design descriptions and practical chair design aspects. The results and comparison of this part of the data will be detailed in Section~\ref{sec: result}. Our keyword extraction criteria centered around functional descriptions and aesthetic elements, following methodologies akin to those used in GenAI-assisted design process analysis, as discussed in \cite{chen2024designfusion}. To maintain fairness and consistency between two sets of data, the keyword extraction process was carried out independently by two authors. The generated results from different groups were randomly shuffled before extraction to eliminate bias. After individual completion, any disputed portions were collectively discussed, and the results were determined. After keyword extraction, we consolidated homonyms to make the following analysis more scientific. An example of our keyword extraction and homonym consolidation process is detailed in Appendix C in our supplementary materials. Following the consolidation, we quantified the frequency of terms used in the ChatGPT solutions and those in the Red Dot solutions. We then identified and calculated the unique words and shared words between the two datasets, tallying both the number of entries and their cumulative frequencies.

To evaluate the novelty in the design solutions contributed by generative AI compared to human-generated solutions, we define the proportion of novelty ($P_{novelty}$) as the ratio of unique words to the total number of word entries (unique and shared). This ratio is given by the formula:
\begin{equation}
P_{novelty} = \frac{U}{U + S}
\label{eq:novelty_ratio}
\end{equation}
where $U$ denotes the number of unique word items and $S$ denotes the number of shared word items.

\paragraph{\textbf{Image generation analysis}}
In our study, we leveraged the CLIP model for an in-depth analysis of chair images generated by Midjourney and compared these with human-designed chair data. Based on insights from \cite{gandelsman2023interpreting} regarding the CLIP-ViT image encoder, we focused on specific attention heads—Layer 22 Head 1, Layer 22 Head 11, and Layer 23 Head 12—which are linked to semantic roles such as “shape,” “color,” and “texture.” We processed chair images \(I\) through the encoder, extracting embeddings for these attributes from the class token \(z_0\) outputs at the identified heads \(z_0^{l,h}\). Additionally, we captured the general embedding from the last layer output of the class token. For visualization, we used t-SNE \cite{van2008visualizing} to reduce these high \(d\)-dimensional embeddings of all images to a 2D space, which determines the coordinates for visualizing the image distribution.

Specifically, the background of both GenAI images and human images were adjusted to white using object recognition technology to minimize distractions. The visualization results are shown in Figure~\ref{fig_figure_TSNE}.

Additionally, we calculated the pairwise distance between the two datasets. To assess the statistical significance of the differences in diversity between images generated by Midjourney and those recognized in design contests, we employed the Mann-Whitney U test for pairwise distance comparisons. The mean distance is computed using the following formula:

\begin{equation}
\overline{D} = \frac{1}{n(n-1)} \sum_{i=1}^{n} \sum_{j=i+1}^{n} d(x_i, x_j)
\end{equation}

where \(d(x_i, x_j)\) represents the distance between two elements \(x_i\) and \(x_j\) in the dataset, and \(n\) is the number of elements in the dataset.

\section{Findings}
\label{sec: result}
In this section, we analyze the manifestations of GenAI design fixation by dividing our analysis into text generation models and image generation models (RQ1). We summarize these fixation manifestations based on the results of our quantitative analysis of text and image data, as well as the design solutions submitted by the participants. Subsequently, we delve into the feedback provided by participants regarding their recognition of GenAI design fixation (RQ2) and its impact on their ideation processes (RQ3).

\subsection{Manifestations in text generation design fixation}
During the data pre-processing stage of our quantitative analysis (see detailed results in Table~\ref{tab:chair_parts_comparison}), we found that the Red Dot data contained more chair design-structure-related keywords than the ChatGPT solutions (208.57 and 142.71 respectively, normalized per 100 cases). After excluding these chair-structure-related words, we focused on analyzing function- and aesthetic-related keywords. For the Red Dot data, we extracted a total of 398 keyword items with a cumulative frequency of 1,104. In the ChatGPT data, 266 keyword items were identified with a cumulative frequency of 712. Our subsequent analysis concentrated on the frequency of these two sets of keywords.

\begin{table}[ht]
\centering
\caption{Comparison of frequencies of keywords related to chair structural elements in design solutions in Red Dot and ChatGPT, normalized per 100 cases.}
\label{tab:chair_parts_comparison}
\begin{tabularx}{0.5\textwidth}{Xrr}
\toprule
\textbf{Part} & \textbf{Red Dot (per 100)} & \textbf{ChatGPT (per 100)} \\
\midrule
Seat          & \calc{70}{105}             & \calc{44}{96}                   \\
Leg           & \calc{49}{105}             & \calc{10}{96}                   \\
Backrest      & \calc{29}{105}             & \calc{33}{96}                   \\
Frame         & \calc{23}{105}             & \calc{9}{96}                    \\
Base          & \calc{22}{105}             & \calc{11}{96}                   \\
Armrest       & \calc{15}{105}             & \calc{14}{96}                   \\
Cushion       & \calc{7}{105}              & \calc{9}{96}                    \\
Footrest      & \calc{2}{105}              & \calc{5}{96}                    \\
Headrest      & \calc{2}{105}              & \calc{2}{96}                    \\
\midrule
\textbf{Total} & \textbf{\calc{219}{105}}  & \textbf{\calc{137}{96}}         \\
\bottomrule
\end{tabularx}
\end{table}

\begin{figure*}[htp]
    \centering
    \includegraphics[width=0.95\textwidth]{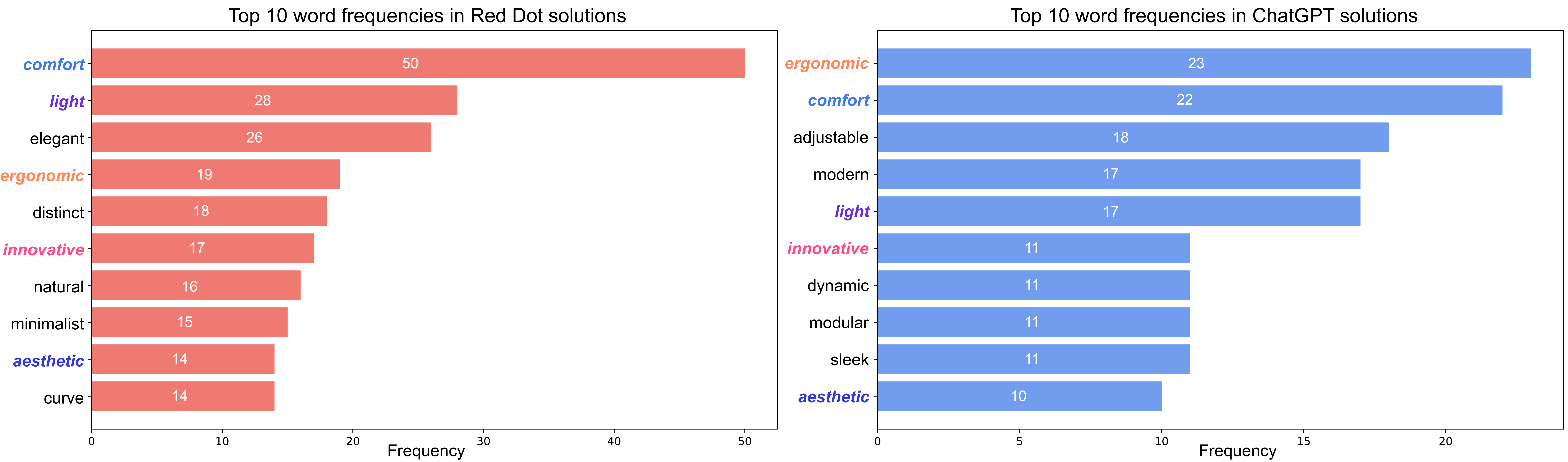}
    \caption{Comparison of the top 10 most frequent word stems in design solutions generated by ChatGPT and those recognized in Red Dot award-winning designs. Common terms across both datasets are highlighted and include "comfort", "light", "ergonomic", "innovative", and "aesthetic".}
    \label{fig_word_frequency}
    \Description{}
\end{figure*}

Figure~\ref{fig_word_frequency} displays the top 10 most frequent word derived from both the ChatGPT-generated and Red Dot design solutions. Notably, both datasets share five common words among their top 10 frequencies: “comfort”, “light”, “ergonomic”, “innovative”, and “aesthetic”. This overlap highlights recurrent keywords in both human and GenAI-generated design solutions.

We also calculated the proportion of unique items in the two set of data. The item counts and frequencies of unique and shared words for Red Dot and ChatGPT solutions are listed in Table~\ref{tab:word_frequencies}. During assessment, we focus on the counts of distinct and shared word entries rather than their total frequencies. This approach highlights the diversity of terms used rather than the volume of usage. Applying formula \ref{eq:novelty_ratio}, the proportions of novelty for both the GenAI-generated data (ChatGPT) and human-generated data (Red Dot) are calculated ($P_{novelty}^{ChatGPT}$ = 67.4\%, $P_{novelty}^{Red Dot}$ = 78.2\%). This demonstrates that the human-generated data exhibits a higher proportion of unique terms, implying a greater diversity in creative outputs compared to the ChatGPT-generated data in our experiment. 

\begin{table}[ht]
\centering
\caption{Item counts and frequencies of unique and shared words for Red Dot and ChatGPT solutions.}
\label{tab:word_frequencies}
\begin{tabular}{lrrrr}
\hline
& \multicolumn{2}{c}{Red Dot} & \multicolumn{2}{c}{ChatGPT} \\
\cline{2-5}
& Item Counts & Frequency & Item Counts & Frequency \\
\hline
Unique Words & 312 & 557 & 180 & 297 \\
Shared Words & 86 & 547 & 86 & 361 \\
\hline
\end{tabular}
\end{table}

Based on the quantitative analysis above and the manual analysis by two of the researchers, we categorized the design fixation in text generation models into four dimensions: \textit{Descriptive statements}, \textit{Repetitive theme}, \textit{Limited contextual variation}, \textit{Susceptibility to high-frequency words}. A summary of these dimensions, along with detailed explanations and examples, is provided in Table~\ref{tab:manifestations_text}.

\textbf{Descriptive Statements.} This refers to the phenomenon where the generated content often appears creative at first glance but lacks depth or clear operational mechanism, primarily focusing on wordplay rather than functional innovation. For instance, when P2 prompted ChatGPT to generate chair designs based on a puzzle structure, the solutions ChatGPT produced (e.g., \textit{An office chair designed with interlocking pieces that can be assembled or disassembled, reflecting the puzzle's structure.}) were mainly superficial descriptions that did not effectively incorporate the design characteristics or structural elements essential to chair design. \textit{P2: "It would be prudent to explore the potential of utilizing assembled or disassembled features, but how remains a question."} This manifestation aligns with previously discussed challenges concerning the limited reasoning capabilities of LLMs \cite{wu2022ai}. These challenges highlight that LLMs often only grasp the form of language without truly understanding the underlying meaning \cite{bender2020climbing}.

\textbf{Repetitive Theme.} This dimension refers to the recurrence of similar concepts or thematic elements across different outputs, where the model consistently generates alike ideas regardless of the specific input details. For instance, when P4 requested a design solution themed around "cloud," aiming for a cozy and soft atmosphere, ChatGPT continued to produce descriptions typical of modern tech offices, characterized by sleek and minimalist designs. This indicates that the model fixated on the common association of "cloud" with technology and cloud computing, neglecting the participant's intended theme of comfort and softness. Such fixation limits the diversity of ideas and can hinder the creative exploration of alternative interpretations of a theme.

\textbf{Limited Contextual Variation.} This dimension highlights the model's difficulty in adapting its responses to different contextual cues, leading to similar outputs even when the context changes significantly. For example, when P3 requested design solutions for a pet area suitable for working from home, ChatGPT continued to provide suggestions appropriate for office chair design, fixating on language and concepts typically used in that context rather than adapting to the new scenario. This suggests that the model is not effectively incorporating new contextual information into its responses, resulting in a narrow range of solutions that may not align with the user's specific needs or the unique aspects of the design challenge.

\textbf{Susceptibility to High-Frequency Words.} This reflects how GenAI often disproportionately favors words or phrases that frequently appear in its training data. This bias may lead to creative solutions that feel incongruent or limited in their inventiveness. For instance, when P6 instructed ChatGPT to generate a creative design based on bamboo, the output was unexpectedly anchored to high-frequency terms "ergonomics": "A bamboo chair designed with an ergonomic structure to enhance comfort and posture support". This response highlights the model's tendency to revert to common concepts like ergonomics, regardless of the specific creative context provided. The comparison of word frequencies between ChatGPT and Red Dot projects, as illustrated in Figure~\ref{fig_word_frequency}, further reveals this disparity. The top words in ChatGPT outputs, such as "ergonomic", "sustainable", and "modular", contrast sharply with more diverse terms found in human-generated, award-winning design descriptions. 

\begin{table}[ht]
\centering
\caption{Manifestations of design fixation in text generation by GenAI.}
\label{tab:manifestations_text}
\begin{tabular}
{p{0.2\textwidth}p{0.75\textwidth}} 
\toprule
\textbf{Dimensions} & \textbf{Explanations}\\ 
\midrule
\midrule
Descriptive statements & The generated text appears creative at first glance but lacks depth or practical application, focusing more on word play rather than functional innovation.\\
\midrule
Repetitive theme & The recurrence of concepts or thematic elements across different outputs, repeatedly churning out similar ideas or concepts regardless of the input specifics.\\
\midrule
Limited contextual variation & The model might struggle to adapt its responses to different contextual cues, leading to similar outputs across varying contextual requirement.\\
\midrule
Dependency on high-frequency words & GenAI disproportionately favors words or phrases that appear frequently in its training data, potentially overshadowing more relevant but less frequent terms.\\
\bottomrule
\end{tabular}
\end{table}


\subsection{Manifestations in image generation model design fixation.}
\begin{figure*}[htp]
    \centering
    \includegraphics[width=0.95\textwidth]{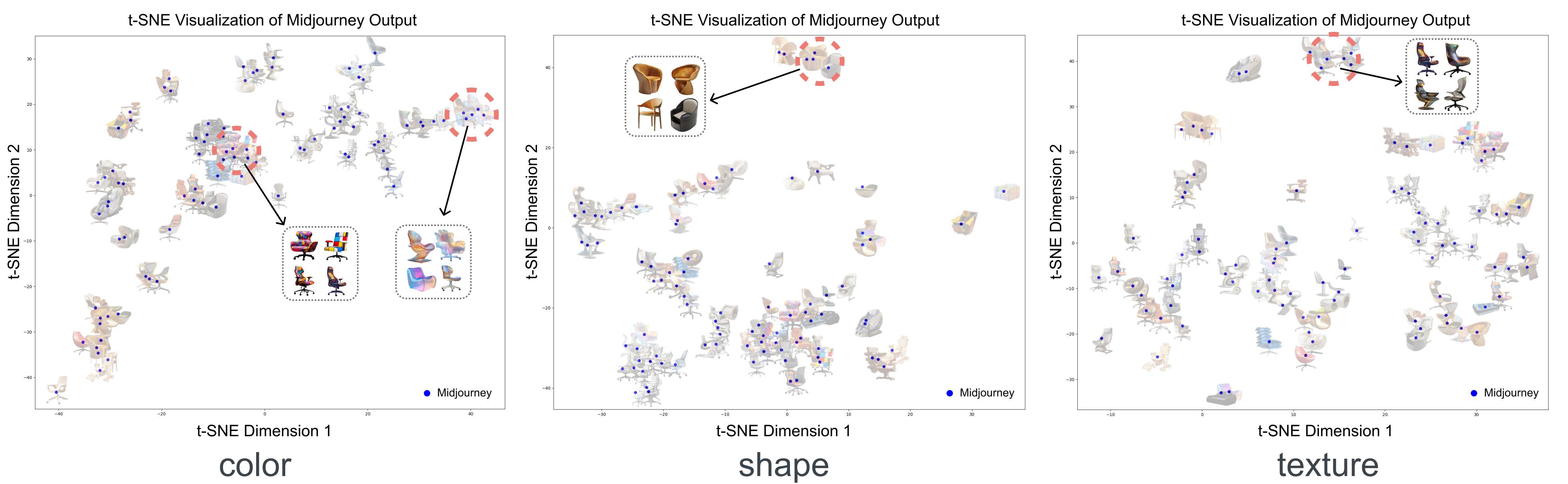}
    \caption{Visualization of t-SNE dimensionality reduction applied to the embeddings from Midjourney-generated chair images.}
    \label{fig_figure_TSNE}
    \Description{}
\end{figure*}

Table~\ref{tab:pairwise_distance} presents the pairwise distance analysis results between Red Dot and Midjourney image datasets, which reveals significant differences across four key design attributes: global features, shape, color, and texture. Notably, the global feature distance is higher for Red Dot solutions (\(M = 16.2509, SD = 3.2638\)) compared to Midjourney-generated solutions (\(M = 16.0503, SD = 4.3814\), \(p = 0.0238^*\)). Additionally, for all other categories—shape, color, and texture—the pairwise distances are consistently larger in the Red Dot solutions compared to Midjourney, with significant \(p\)-values indicating these differences (\(p < 0.01\)). Besides, the result of exploratory analysis of the image attributes is shown in Figure~\ref{fig_figure_TSNE}, using the visualization of t-SNE dimensionality reduction applied to the embeddings from Midjourney-generated chair images. We marked several representative clustering to facilitate further analysis of the GenAI design fixation it represents.

\begin{table*}[ht]
    \centering
    \caption{Pairwise distance analysis between Red Dot and Midjourney image datasets.}
    \label{tab:pairwise_distance}
    \renewcommand{\arraystretch}{0.8}  
    \begin{threeparttable}
        \begin{tabularx}{0.7\textwidth}{l *{2}{>{\centering\arraybackslash}X} *{2}{>{\centering\arraybackslash}X} c}
            \toprule
            Category & \multicolumn{2}{c}{Red Dot} & \multicolumn{2}{c}{Midjourney} & p-value \\
            \cmidrule(lr){2-3} \cmidrule(lr){4-5}
            & Mean & SD & Mean & SD \\
            \midrule
            Global   & \textbf{16.2509} & 3.2638 & 16.0503 & 4.3814 & 0.0238\tnote{*} \\
            Shape    & \textbf{0.0700} & 0.0003 & 0.0592 & 0.0003 & 0.0000\tnote{**} \\
            Color    & \textbf{0.0821} & 0.0003 & 0.0685 & 0.0002 & 0.0000\tnote{**} \\
            Texture  & \textbf{0.0882} & 0.0004 & 0.0862 & 0.0006 & 0.0000\tnote{**} \\
            \bottomrule
        \end{tabularx}
        \begin{tablenotes}
    \item[] \textbf{Notes:}
      \item 1. Bold font represents bigger pairwise distance.    
    \item 2. ** denotes $p < 0.01$ and * denotes $p < 0.05$.
    \end{tablenotes}
    \end{threeparttable}
\end{table*}

\begin{figure*}[htp]
    \centering
    \includegraphics[width=0.95\textwidth]{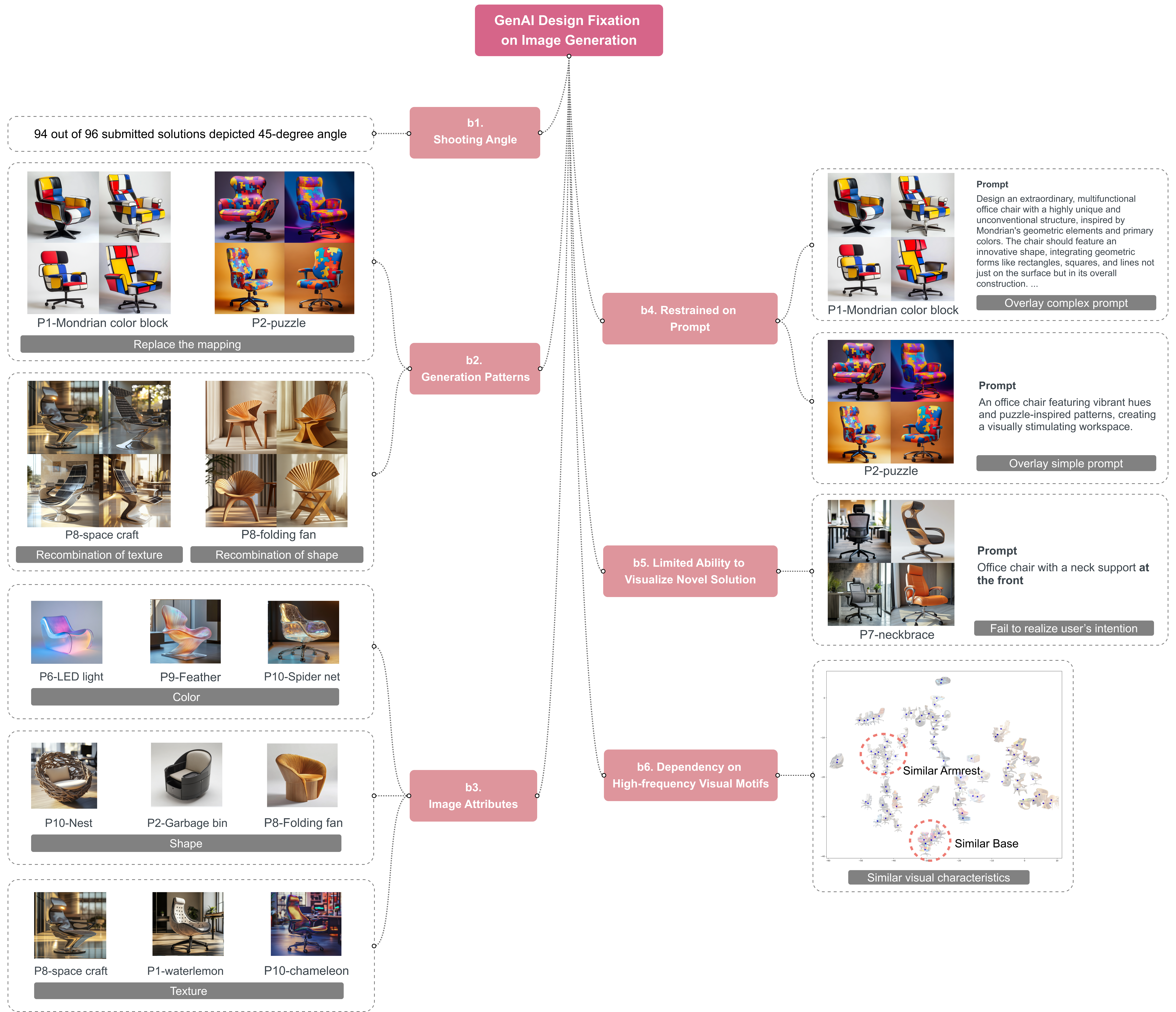}
    \caption{Manifestations of design fixation on image generation models from our experiment displayed on the left.}
    \vspace{-0.05in}
    \label{fig_image_manifestion}
    \Description{}
\end{figure*}

Based on the analysis results above and the manual analysis for users' submitted solutions, we categorized the design fixation in image generation models into seven dimensions: \textit{Shooting angle}, \textit{Image generation patterns}, \textit{Image attributes}, \textit{Restrained on prompt}, \textit{Limited ability to visualize novel solution descriptions}, and \textit{Dependency on high-frequency visual motifs}. A summary of these dimensions, along with detailed explanations and examples, is provided in Table~\ref{tab:manifestations_image} with corresponding examples shown in Figure~\ref{fig_image_manifestion}.

\textbf{Shooting angle.} This indicates that generated images typically showcase a limited range of angles or viewpoints, predominantly those most commonly represented in the training data. Specifically, in our study, out of 96 user-submitted chair design solutions, only 2 depicted front views; the remainder predominantly featured 45-degree angles.

\textbf{Image generation patterns.} Our analysis revealed two distinct image generation patterns in the Midjourney outputs: surface mapping replacement of common chairs and the recombination of textures and shapes. As shown in Figure~\ref{fig_image_manifestion}-b2, these patterns represent GenAI’s tendency to focus on aesthetic alterations rather than overall structure consideration. For instance, P1 and P2’s designs replace the conventional ergonomic chair surfaces with Mondrian color blocks and puzzle-themed patterns, adding a superficial layer to the standard chair form. On the other hand, P8’s design exemplifies texture recombination by integrating spacecraft materials into the chair’s surface, which visually disrupts its traditional appearance without modifying its functional structure. In another instance, the expanded form of a folding fan was used to replace the traditional chair back, serving as a purely visual recombination that does not consider the design’s practicality or structural feasibility.

\textbf{Image attributes.} This observation comes from the clustering phenomenon observed in the t-SNE reduction of color, shape, and texture attributes. As shown in Figure~\ref{fig_image_manifestion}-b3 (consistent with the marked areas in Figure~\ref{fig_figure_TSNE}), despite different prompts like P6-LED light, P9-feather, and P10-spider net, which have varying additives and requirements, the output images exhibit similar color patterns. This phenomenon is also evident in the dimensionality reduction results for shape and texture.

\textbf{Restrained on prompt.} As exemplified in Figure~\ref{fig_image_manifestion}-b4, whether the prompts are complex or overly simplistic, the output frequently converges on a conventional office chair design. This indicates a limitation in the model’s response range to diverse prompt complexities, often defaulting to standard, familiar outputs regardless of the creative potential of the input.

\textbf{Limited ability to visualize novel solution descriptions.} As demonstrated in Figure~\ref{fig_image_manifestion}-b5, when P7 describes a chair featuring a neck brace—a component possibly absent from the training data—the model fails to accurately visualize this feature in the generated image. This highlights the model’s struggle to produce images that depict novel solutions or concepts not represented within its training dataset.

\textbf{Dependency on high-frequency visual motifs.} Image generation models often over-rely on visual motifs that appear frequently within their training datasets. For instance, in the generation of chair images, TNSE analysis of global features in Figure~\ref{fig_image_manifestion}-b6 reveals that the model consistently produces similar designs for armrests and bases. This pattern reflects the prevalent styles learned from the training data and indicates a limited exploration beyond these familiar designs, despite the potential for a broader array of creative interpretations.

\begin{table}[ht]
\centering
\caption{Manifestations of design fixation in image generation by GenAI.}
\label{tab:manifestations_image}
\begin{tabular}
{p{0.2\textwidth}p{0.75\textwidth}}
\toprule
\textbf{Dimensions} & \textbf{Explanations}\\ 
\midrule
\midrule
Shooting angle & Generated images primarily feature a limited set of angles or viewpoint, often those most commonly seen in the training data. \\
\midrule
Image generation patterns & Images are mostly produced through replacing the mapping of the outer surface of the common chair, or recombination of textures and shapes, lacking more creative embodiment methods. \\
\midrule
Image attributes & Marked homogeneity in attributes such as color, shape, and texture. The model tends to replicate these familiar attributes regardless of the input prompt's requirements. \\
\midrule
Restrained on prompt & Complex or overly simple prompts generally result in obtaining a conventional office chair. 
\newline
Creative combinations often require users to provide hints in the prompt. \\
\midrule
Limited ability to visualize novel solution descriptions & The model struggles to generate images that depict solutions or concepts not present within its training data. \\
\midrule
Dependency on high-frequency visual motifs &  Image generation models tend to overuse visual motifs that are frequently encountered within their training datasets. \\
\bottomrule
\end{tabular}
\end{table}

\subsection{Participants' feedback}
\label{subsection_Participants'_feedback}

In this section, we discuss participants' feedback on their experience interacting with GenAI during the office chair design ideation stage. The feedback is drawn from the interview data as well as the observations made by researchers during the experiment. 

\textbf{Initial ideation.}
Approximately one-third (30\%, n=3) of the participants directly asked ChatGPT for creative ideas related to designing chairs for office settings or requested innovative "additives" during the experiment. Other participants chose to provide their own ideas and concepts to ChatGPT, asking it or GenAI-based Creativity Support Tools (CSTs) to further explore and expand upon these initial inputs. This suggests that the influence of GenAI outputs on human designers' ideas may begin to take effect early in the design process. For more experienced designers, critical thinking is applied to evaluate ChatGPT's suggestions. For instance, P8 remarked during the experiment that many of ChatGPT's creative elements, whether visual or functional, were overly focused on intelligentized or futuristic design features (e.g., adding LED lights). As a result, P8 rejected all of ChatGPT’s proposals and developed his own design ideas instead. In contrast, novice designers may be more inclined to adopt ChatGPT's ideas directly. For example, P6 requested innovative solutions from ChatGPT regarding form and color and subsequently fed these unaltered ideas into Midjourney for further development.\\

\textbf{Tool usage habits and prompt iteration.}
During the co-creation process using ChatGPT and Midjourney, nearly all participants tended to directly copy the product descriptions or Midjourney prompts generated by ChatGPT and input them into Midjourney without modification, and only two participants independently modified the Midjourney prompt provided by ChatGPT during the experiment. Some participants, such as P3, indicated that they paid little attention to the specific content of ChatGPT's product descriptions or Midjourney prompts. Instead, they focused on evaluating whether the final outcomes produced by Midjourney were sufficiently innovative. As a result, these participants often overlooked potential issues in ChatGPT's suggestions, as well as the risk of the solutions lacking creativity and diversity.\\

\textbf{Recognition of GenAI design fixation.}
Overall, when engaging with ChatGPT and Midjourney, seven and eight participants, respectively, responded affirmatively to the interview question (1): "Have you noticed any repetition or similarities in the ideas or designs generated by the tools?" This indicates a recognition of some level of GenAI design fixation during the 30-minute formal experiment involving interactions with GenAI.

Regarding the text generation model, their insights primarily focused on ‘fixation on descriptive statements’ (P2) and ‘susceptibility to high-frequency words’ (P4, P9) as cataloged in Table~\ref{tab:manifestations_text}. For instance, when P2 requested a creative chair design inspired by a fountain from ChatGPT, the response was, \textit{“An office chair designed with layered tiers for enhanced ergonomic support, inspired by the structure of a fountain”.} P2 commented, \textit{“The response from ChatGPT did not integrate well with the chair’s structure; it required further prompting to combine the tiered structure of the fountain with the cushion structure to generate a more specific answer.”} Besides, P9 noted, \textit{“The responses from ChatGPT often contain descriptions like ‘modular,’ ‘adaptive,’ ‘automatically adjustable,’ and ‘sensors,’ which I encountered even when designing chairs based on yoga balls and mechanical structures.”} P4 identified ‘ergonomics’ as a frequently mentioned term.

A greater number of participants noticed fixation phenomena in the image generation model, primarily encompassing aspects listed in Table~\ref{tab:manifestations_image}. These include b2-fixation on image generation patterns, such as replacing the mapping (P1) and recombining shapes regardless of ergonomics (P8); b5-limited ability to visualize novel solutions, exemplified by the inability to integrate a neck brace into the design (P7); and b6-dependency on high-frequency visual motifs, noted by P2 in terms of similar and commonly seen base and armrest structures. The respective cases from participants are displayed in Figure~\ref{fig_image_manifestion}, illustrating these specific manifestations of design fixation within the GenAI output.\\


\textbf{The impact of GenAI design fixation on design ideation.}
Based on insights from the design process and interview data with participants, it is evident that the impact of GenAI design fixation on design ideation \textbf{exhibits individual differences.} More experienced designers not only have a broader and richer understanding of design case libraries but also possess relatively mature design thinking and established processes. These attributes aid them in developing a habit of evaluating GenAI outputs critically. For instance, P5, a GenAI enthusiast and designer with six years of experience, was initially interested in a chair design for a pet-friendly office chair that included “a cushioned area beneath or beside the seat where a small to medium-sized pet can relax.” This proposal initially inspired P5, who had been contemplating designs that accommodate pets beside the chair. However, she continued to ponder the practicality of this design and whether there could be a better solution. Consequently, she input the prompt \textit{“with smooth curves and a built-in scratching post along one side”}, which led to a more optimized solution for human-pet interaction (for a detailed example, see Appendix D in our supplementary materials).

During her response to interview question 3, P5 remarked: \textit{“GenAI can inspire creativity, but the design must serve a purpose and be practical. I assess and further iterate on the designs from a practicality standpoint.”} She also noted, “Novice designers using GenAI might forget the purpose of the design and end up selecting GenAI-generated outcomes, especially for their aesthetic effects. However, many elements in these designs have flaws,” highlighting a gap that less experienced participants often encounter.

\section{Addressing GenAI design fixation in inspiration of computational creativity}
\label{section_solution}

According to the definition of GenAI design fixation established in this study and substantiated through our experimental investigation, this phenomenon may originate from data bias, the architectural constraints of algorithms, and the prompts provided by human users. Such occurrences have the potential to limit the creative capacities of GenAI systems, thereby potentially misleading human designers and, to some extent, obstructing access to higher-quality solutions. Recent research has demonstrated that human designers might be more likely to often exhibit fixation on the outputs generated by GenAI \cite{wadinambiarachchi2024effects}. Consequently, alleviating GenAI design fixation would be instrumental in reducing the homogeneity of GenAI design outputs, which serve as stimuli for human designers, thereby enhancing the diversity and creativity of human-AI co-design solutions.

Despite the technical recommendations provided, we propose adopting methods to alleviate GenAI design fixation. Furthermore, we suggest the HCI community to consider the phenomenon of GenAI design fixation when designing and evaluating creativity support tools based on Generative AI.

\begin{enumerate}
    \item \textbf{Reduce bias in design data:} When exploring the training and application of GenAI models, data bias is a critical issue that can't be ignored. Bias in design data can affect the fairness and accuracy of the model, potentially misleading the design process of designers \cite{zhou2024bias, ferrara2023should}. These emphasize the importance of training artificial intelligence models with diverse and balanced datasets, to ensure that the models can produce more equitable, diverse, and innovative design solutions.
    \item \textbf{Optimize model architecture and objective functions:} To enhance the model's innovative capacity and adaptability while avoiding over-optimization and fixation, it is necessary to optimize the model architecture and objective functions. For text and image generation models, this means exploring new architectures capable of understanding and generating complex relationships and adopting strategies to increase the model's sensitivity to low-frequency data.
\end{enumerate}

\subsection{Creativity support tool design}
In this section, we give some reference strategies regarding methods to mitigate human design fixation, researchers have actively explored various strategies to mitigate design fixation \cite{linsey2010study}. According to the solution categorization proposed by \cite{alipour2018review}, the strategies can be divided into sourece, methods and instructions. In this section, we discuss mitigation methods in inspiration of the two directions in the context of Generative AI design fixation, as well as proposing strategies for interaction design. 

\subsubsection{Choosing the right kind of source}
In human design fixation study, the term ‘source’ refers to the use of previous examples and other resources as references for the solution to the current problem \cite{cai2010extended}. Studies have recommended certain strategies for reducing the effect of fixation in design. Their results have demonstrated that some sources leave enough room for exploration in design \cite{cheng2014new} and have a positive effect on the design outcome \cite{goldschmidt2011avoiding}. 

Analogously, in GenAI design fixation, studies has shown that providing additional, non-routine knowledge bases for GenAI-based CSTs can enhance the creativity and diversity of GenAI outputs. For instance, integrating specialized biological knowledge databases into GenAI systems \cite{zhu2023biologically} has been shown to expand the scope of design possibilities, encouraging the exploration of innovative solutions that might not emerge from conventional datasets \cite{kang2024biospark}. Similarly, supplying databases from specific fields of knowledge, such as materials science, cultural studies, or even niche areas of art and literature \cite{wang2024promptcharm}, can guide GenAI to venture into new, uncharted territories of design. These enriched knowledge bases serve as catalysts, prompting GenAI to generate more varied and inventive outputs.

\subsubsection{Instructions and methods}
In addition to choosing the right kind of source, studies on human design fixation has also highlighted the effectiveness of specific design instructions or systematic methods in order to help designers overcome design fixation. Several notable strategies include group working \cite{youmans2011effects}, employing alternative representations of the problem \cite{linsey2010study}, developing instructional mapping rules \cite{cheong2013using}, and utilizing the design-by-analogy method \cite{linsey2012design}. These approaches have been shown to diversify thinking and expand the range of solutions considered during the design process.

These methods can also be transferred to the field of GenAI to alleviate its design fixation. As previous empirical studies pointed out GenAI's tendency to produce surface-level or basic information, particularly in the context requiring in-depth research or exploration \cite{kobiella2024if}. For example, by employing the method of Multi-Agent Collaboration, we can endow GenAI with critical thinking and iterative capabilities. Adopting a Human-AI Collaboration system \cite{lee2024conversational} allows us to combine human creativity with AI capabilities. For instance, use GenAI to generate initial ideas, but have humans refine, combine, or expand upon these ideas. This hybrid workflow can reduce the risk of becoming fixated on AI-generated outputs. It can also guide GenAI to make broader associations, including proposing alternative representations of the problem and using analogies.

\subsubsection{Interaction design}
In addressing the issue of design fixation in GenAI, interactive measures and the presentation of outputs play a crucial role. Firstly, by providing inspiring example solutions, users can be guided to think critically rather than being given direct answers. This can be achieved through user interfaces or workflows that stimulate creative thinking. 

Additionally, explicitly pointing out flaws in examples and providing instructions to avoid problematic elements can help users identify and circumvent potential design pitfalls. 

Secondly, tool customization allows users to adjust the randomness of AI-generated outputs, explore different styles, and set specific goals, thereby preventing repetitive and predictable results. \cite{liu2022design} 


\subsubsection{Improve prompt engineering}
As a consensus in HCI community, GenAI are sensitive to input prompts \cite{wu2022ai}, which is also one of the causes of GenAI design fixation in our study. As observed in our experiment and in line with other HCI study conclusion , novice designers often face challenges in crafting effective prompts for GenAI \cite{zamfirescu2023johnny}, with issues arising from prompts that are either too complex or too simple. Common issues include prompts that are overly complex or overly simplistic. This challenge has also been documented in previous studies focusing on creativity support for designers \cite{chen2024designfusion, liu2022design}. From the perspective of GenAI design fixation, these insights underscore the critical role of prompt engineering in fostering more diverse and innovative design outputs.

Effective prompt engineering requires a balance between clarity and flexibility, enabling GenAI to explore a broad range of creative possibilities. Prompts should guide the GenAI with enough specificity to maintain relevance while allowing room for creative interpretation. For example, specifying a theme or mood can direct the AI while still permitting innovative variations.
though the GenAI's stochastic nature to explore possibilities could yield creative ideas sometimes, the trial-and-error process is time-consuming.

Iterative refinement is also crucial \cite{mahdavi2024ai}. By evaluating the outputs from initial prompts, designers can adjust subsequent prompts to better harness GenAI’s creative potential. This process involves a systematic approach where experimenting with various prompts can help understand how different prompt attributes affect the creativity of the outputs. Ultimately, the goal is to develop a toolkit for designers that supports the consistent elicitation of high-quality creative outputs from GenAI systems, enhancing the technology’s role in the creative process.


\subsection{Education}
\subsubsection{Understanding the phenomenon of GenAI design fixation}
Howard et al.'s study \cite{howard2013overcoming} reflects that educating students in the phenomenon and effects of fixation enables them to effectively devise their own strategies to avoid or overcome fixation. So the first step in addressing design fixation within GenAI involves ensuring that designers comprehend the nature and implications of this phenomenon. Our work in align with \cite{anderson2024homogenization}'s findings that users given a sense of what the GenAI tend to suggest in similar contexts could help mitigate homogenization effects. In the context of GenAI, design fixation specifically pertains to the GenAI's tendency to generate outputs that adhere too closely to learned patterns and examples, thereby stifling novel and diverse designs.

\subsubsection{Users need to critically evaluate AI-generated content}
To mitigate the effects of GenAI-induced design fixation, it is crucial to train users to critically evaluate AI-generated content rather than accepting it passively. The importance of guiding designers reflection on generated designs is also proposed in other HCI research, such as \cite{gmeiner2023exploring}. This training should encompass an understanding of the underlying algorithms to some extent, enabling users to recognize the inherent limitations and bias of the GenAI. By fostering a critical mindset, asking questions like \textit{“Is there a better design approach?”} and \textit{“What limitations might the GenAI-generated results have?”}, designers can more effectively assess the suitability and originality of AI outputs. This approach ensures that these tools serve as a starting point for further creative development rather than as definitive solutions.



\subsection{Evaluation metrics for GenAI-based CSTs research}

To effectively evaluate GenAI-based creativity support tools (CSTs), it is essential to incorporate design fixation as a critical standard. Design fixation, the tendency to become overly influenced by existing examples or solutions, can significantly hinder creativity and innovation. Therefore, any assessment framework for GenAI-based CSTs must rigorously examine the extent to which these tools either mitigate or exacerbate design fixation.

\subsubsection{Positioning of GenAI CST Tools}

GenAI CST tools should be positioned not merely as instruments for increasing the efficiency and quantity of design outputs but as catalysts for enhancing designers' creative and innovative capacities. The primary goal of these tools should be to stimulate designers' thinking, helping them generate more creative, innovative, and groundbreaking ideas and inspirations, which contrasts with a narrow focus on efficiency and productivity.

A key consideration is to avoid overly programmatic workflows that may strengthen design fixation. While structured processes can streamline tasks, they might also limit the creative potential of designers by promoting adherence to predefined patterns and solutions. The development of GenAI-based CST tools should not solely focus on maximizing the AI's capabilities. Given the inherent risk of fixation within GenAI, it is crucial to design these tools in a way that incorporates significant designer participation. Emphasis should be placed on how these tools can engage the designer's subjective agency and imagination, thus fostering a more collaborative and dynamic creative process.

\subsubsection{Comprehensive evaluation of GenAI CSTs}

The evaluation of GenAI CST tools should extend beyond traditional usability tests and the resolution of issues identified in formative studies or user studies. While usability and problem-solving are important metrics, they do not fully capture the tools' impact on designers' cognitive and creative processes. Research outcomes should include rigorous assessments of how well these tools stimulate designers' cognitive engagement and creative thinking. This involves evaluating whether the CST tools genuinely enhance designers' ideation processes and their ability to conceptualize innovative solutions. Evaluations should also measure the impact of CST tools on the overall vibrancy and originality of designers' thought processes. This can include metrics such as the diversity and novelty of ideas generated, the ability to break away from conventional patterns, and the overall enhancement of creative problem-solving skills.

By integrating these comprehensive evaluation criteria, research on GenAI-based CSTs can ensure that these technologies not only address practical usability concerns but also significantly contribute to the advancement of creative design practices. This holistic approach will help in developing tools that truly empower designers, fostering an environment where human creativity and AI capabilities synergistically drive innovation while effectively mitigating the risk of design fixation.
\section{Discussions}

\subsection{Distinctions about GenAI design fixation and other terms related to GenAI limitations}
In this section, we clarify the distinctions between GenAI design fixation and other terms often discussed in HCI associated with limitations within the GenAI research domain. We believe that understanding these differences would be beneficial for effectively addressing the specific challenges that arise in the development and application of Generative AI technologies


\subsubsection{GenAI error}
GenAI error refers to the phenomenon where GenAI produces content that contradicts factual information \cite{liu2024smart}. This issue is particularly pertinent in contexts such as code generation \cite{ebert2023generative}, where bugs or unwanted code fragments may be introduced, or in text generation, where factual inaccuracies can occur. In the domain of AI-generated text, this can manifest in various applications including reasoning, translation, summarization and paraphrasing, and content evaluation, as well as in the creation of creative content. The presence of GenAI errors in these contexts can lead to misleading outcomes, which not only diminish the reliability and accuracy of the generated content but also potentially misguide users. However, AI “error” also plays a positive role in the early ideation stage, supporting designers to find association through four forms of creativity named CETR \cite{liu2024smart}.

\subsubsection{GenAI hallucination} 
GenAI hallucination refers to the phenomenon where Generative AI produces content that is nonsensical or unfaithful to the provided source material \cite{ji2023survey}. This issue is particularly salient in several key applications, including abstractive summarization \cite{pagnoni2021understanding}, data-to-text generation \cite{wiseman2017challenges}, dialogue systems \cite{li2020slot, santhanam2021rome}, generative question answering (GQA) \cite{nguyen2016ms}, translation tasks \cite{zhou2020detecting}, and computer vision \cite{zhou2023analyzing}. In these contexts, hallucinated content can lead to misinterpretations and inaccuracies, undermining the reliability and trustworthiness of the AI-generated output. However, in the realm of creative content generation, while hallucinations can introduce misleading elements, they can also paradoxically stimulate designers' creativity by providing unexpected and novel ideas. This dual impact underscores the complexity of addressing GenAI hallucinations: while it is crucial to minimize their occurrence to ensure the fidelity and accuracy of AI-generated content, it is also important to recognize their potential to inspire innovative thinking and creative exploration.

\subsubsection{GenAI bias}
GenAI hallucination refers to the presence of systematic misrepresentations, attribution errors, or factual distortions that result in favoring certain groups or ideas, perpetuating stereotypes, or making incorrect assumptions based on learned patterns. This phenomenon is particularly relevant in the contexts of image generation \cite{zhou2024bias} and text generation \cite{ferrara2023should}. In these domains, hallucinations can manifest as biased or inaccurate representations that reinforce existing societal bias and stereotypes. By highlighting and confronting these bias, designers may be inspired to develop more inclusive and equitable creative solutions.

\subsubsection{GenAI creativity}
GenAI creativity refers to the ability of GenAI systems to produce novel and innovative outputs in creative domains such as art, design, music, and writing. These systems are increasingly used to generate unique artworks, pushing the boundaries of traditional processes by introducing fresh perspectives and solutions. However, despite these advancements, the outputs of GenAI, particularly in writing, may not always match the depth and expertise of human specialists. While large language models can produce coherent text, they often lack the nuanced quality that expert human writers possess \cite{chakrabarty2024art}. The impact of GenAI creativity is dual: it democratizes access to creative tools, fostering a more diverse creative ecosystem, but also highlights the irreplaceable value of human expertise. Thus, GenAI should be seen as a complement to human creativity, enhancing rather than replacing it.


\subsection{Pros and cons of GenAI design fixation for human creativity}

With the prevailing trend in applying GenAI into huamn creativity support, and given that recent studies indicate designers are more prone to fixating on GenAI outputs \cite{wadinambiarachchi2024effects}, it is crucial to discuss the impact of GenAI design fixation on GenAI outputs, the ideation process of designers, and the final design outcomes. Similar to the perspectives on design fixation within the design community \cite{smith1993constraining, bilalic2008good}, we consider the effects of GenAI design fixation to be a double-edged sword.

For the advantages of GenAI design fixation:
\begin{enumerate}
    \item \textbf{Efficiency and Speed: }Similar to human designers who rely on familiar patterns or established solutions, GenAI can use design fixation to quickly generate outputs based on existing data and learned patterns. This can significantly reduce the time required to produce design variations, especially in contexts where rapid prototyping or iteration is necessary.
    \item \textbf{Consistency and Reliability: }GenAI design fixation can help maintain consistency across a series of designs by adhering to specific, proven design elements. This is particularly useful in fields where uniformity is critical, such as branding, where a cohesive visual identity must be maintained.
    \item \textbf{Leveraging Established Best Practices: }By fixating on successful designs or widely accepted principles, GenAI can ensure that its outputs are grounded in established best practices. Besides, studies have shown that access to GenAI ideas may increase the creativity and quality of the outputs,especially among less creative or novice users \cite{doshi2024generative}.
    \item \textbf{Cost-Effectiveness: }By reusing and adapting existing design elements, GenAI can reduce the resources needed for innovation, thereby saving costs related to experimentation, testing, and validation. This is similar to how repeating past solutions in human design can save time, money, and effort.
\end{enumerate}

For the disadvantage of GenAI design fixation:
\begin{enumerate}
    \item \textbf{Limited creativity and innovation: }One of the major drawbacks of design fixation, whether human or AI-driven, is the potential stifling of creativity. When GenAI overly relies on existing patterns or past data, it may struggle to generate truly novel or innovative designs, limiting the scope of creative exploration.
    \item \textbf{Reinforcement of bias: }If the data GenAI is trained on reflects certain bias or limitations, fixation on these patterns can perpetuate and even amplify these bias. This can lead to a lack of diversity in design outputs and may inadvertently reinforce stereotypes or outdated concepts. 
    \item \textbf{Adaptability challenges: }In rapidly changing fields, adherence to past designs may result in outputs that are not well-adapted to new trends or emerging user needs. GenAI design fixation might make it difficult for AI to pivot and adapt to new paradigms, leading to designs that feel outdated or out of touch.
    \item \textbf{Box human creativity in the long run: }As a recent empirical study find that the use of ChatGPT in creative tasks resulted in increasing homogenized contents, and this homogenization effect persisted even when ChatGPT was absence, which highlights the challenge of boxing human creative capability in the long run \cite{liu2024chatgpt}. Constantly relying on familiar solutions can lead to design outputs that are safe but unremarkable. This “safe” approach might prevent GenAI from pushing the boundaries and exploring more groundbreaking or avant-garde design possibilities, which could limit its usefulness in highly creative or forward-thinking industries.
\end{enumerate}

In summary, GenAI design fixation is not always counterproductive; rather, it can promote a more focused and thorough exploration of viable solutions, which indicate that the impact of GenAI design fixation can vary greatly depending on the context, the specific demands of the project, and the creative framework employed by the designers. The perspective of GenAI design fixation may bring new insights to balance design efficiency and design innovation in GenAI-based co-creation process.




\subsection{Limitations and future directions}
While not exhaustive, this study marks the commencement of proposing a theoretical framework for GenAI design fixation and explores its manifestations through empirical investigation. Our research employs two representative generative models (GPT-4o and Midjourney); however, in practice, different GenAI models may produce varying results in response to the same prompts. This potential variability was not explored in our study, representing a limitation of our current methodology. Moreover, by specifically targeting novice designers as participants, this study inherently narrows the scope of feedback regarding the impact of GenAI design fixation on the ideation process. 


In this study, we conducted an experimental investigation and documented the manifestations of design fixation within the context of product design, aligning with established precedents in human design fixation research \cite{crilly2017next}. Future studies should consider broadening the research scope to other areas of creativity support. We particularly recommend further investigations in the fields of creative industries, product design, and conversational agents. The manifestations of GenAI design fixation might vary across different task domains, yet they should be compatible within our proposed framework for GenAI design fixation. Moreover, future work could involve conducting literature reviews through the lens of GenAI design fixation, exemplified by \cite{chakrabarty2024art}, which demonstrates the utility of leveraging comprehensive surveys of Creative Support Tools (CSTs).


\section{Conclusion}
This research has clearly and preliminarily demonstrate the existence of GenAI design fixation. The interest in this lens stems from the desire to understand more deeply both the nature of GenAI-assisted creativity support process as well as to propose potential improvements to directions by which GenAI-based creativity process can be improved. The research has raised far more questions than it has answered but has clearly opened up an interesting avenue of investigation. One very important aspect of this research is its clear boundary-spanning character, combing the highly technological discipline of artificial intelligence and a central character of the design sciences. This transfer is a clear demonstration of the multi-disciplinary character of HCI research. The fundamental differences between human design design fixation and GenAI design fixation are great, and the differences between other GenAI imperfection terms. We hope that this research could help other researchers recognize the behaviors and limitations of GenAI in creativity support area.

\bibliographystyle{ACM-Reference-Format}
\bibliography{sections/Reference}

\end{document}